\documentstyle[11pt,aaspp4,epsf]{article}
\def\edth{\;\raise1.0pt\hbox{$'$}\hskip-6pt\partial\;}
\def\baredth{\;\overline{\raise1.0pt\hbox{$'$}\hskip-6pt
\partial}\;}
\def\gsim{~\rlap{$>$}{\lower 1.0ex\hbox{$\sim$}}}
\def\lsim{~\rlap{$<$}{\lower 1.0ex\hbox{$\sim$}}}
\def\HII{H{\sc ii}}
\def\HI{H{\sc i}}
\def\d{{\rm d}}
\begin{document}
\title{ Polarization of the Cosmic Microwave Background from 
Non-Uniform Reionization } 
\author{Guo-Chin Liu$^{1,2,7}$, Naoshi Sugiyama$^{2}$, Andrew J. Benson
$^{3,4}$, C. G. Lacey$^{5}$ and Adi Nusser$^6$}
\bigskip
\affil{$\ ^{1}$Department of Physics, Kyoto University, Kyoto
606-8502,~Japan}
\affil{$\ ^{2}$Division of Theoretical Astrophysics, National
Astronomical Observatory Japan, Mitaka, 
181-8588,~Japan}
\affil{$\ ^{3}$California Institute of Technology, MC 105-24, Pasadena, CA 91125, U.S.A.}
\affil{$\ ^{4}$Department of Physics, University of Durham, UK}
\affil{$\ ^5$SISSA, via Beirut 2-4, 34014 Trieste, Italy}
\affil{$\ ^6$The Physics Department, The Technion-Israel Institute of
Technology, Technion City, Haifa 32000, Israel}
\affil{$\ ^7$Academia Sinica Institute of Astronomy and Astrophysics,
Taipei 106, Taiwan}
\begin{abstract}
We study the polarization anisotropy in the Cosmic Microwave
Background (CMB) resulting from patchy reionization of the IGM by
stars in galaxies.  It is well known that the polarization
of the CMB is very sensitive to the details of reionization, including
the reionization epoch and the density fluctuations in the ionized
gas. We calculate the effects of reionization by combining a
semi-analytic model of galaxy formation, which predicts the redshifts
and luminosities of the ionizing sources, with a high resolution
N-body simulation, to predict the spatial distribution of the
ionized gas. The models predict reionization at redshifts $z \sim
5-10$, with electron scattering optical depths due to reionization
$\sim 0.014-0.05$. We find that reionization generates a peak in the
polarization spectrum with amplitude $ 0.05 \sim 0.15 \mu$K at large
angular scales ($l\sim 3$).
%
The position of this peak reveals the size of the horizon at
reionization, whilst its amplitude is a measure of the optical depth
to reionization. On small scales ($l \gsim 6000$), reionization
produces a second-order polarization signal due to the coupling of
fluctuations in the free electron density with the quadrupole moment
of the temperature anisotropy.
Careful study reveals that this coupling generates equal second-order
polarization power spectra for the electric and magnetic modes, with
amplitude  $\sim 10$nK. This amplitude depends strongly on the
total baryon density $\Omega_{\rm b}$ and on the spatial correlations
of the free electron density, and weakly on the
fraction $f_{\rm esc}$ of ionizing photons able to escape their source
galaxy.
The
first- and second-order signals are therefore sensitive to 
different details of how the 
reionization occurred.  Detection of these signals will place
important constraints on the reionization history of the Universe.
\end{abstract}

\section{INTRODUCTION}

The secondary anisotropies in the Cosmic Microwave Background (CMB)
provide a laboratory for the study of the epoch of reionization in the
Universe.  With the rapid improvement in CMB experimental sensitivity
and resolution, it is timely to discuss the signals introduced by the
process of reionization, which are at the $\mu$K level or below.  In
this paper, we concentrate on the CMB {\em polarization} in models
with inhomogeneous reionization. It is well known that the primordial
CMB polarization is generated at the recombination epoch through
Thomson scattering of the quadrupole of the temperature anisotropy.
The same mechanism operates at the reionization epoch and again
distorts the shape of the polarization power spectrum.  It has been
pointed out by Ng \& Ng (1996) and Zaldarriaga (1997) that this leads
to the polarization anisotropy being suppressed on small scales but
enhanced on large scales.
The degree of suppression and enhancement typically depend on the
optical depth to the recombination epoch produced by
reionization. Although the amplitude of the 
polarization anisotropy is estimated to be much smaller than that in
the temperature, it is advantageous to consider the polarization
signal since it is generated when photons and electrons scatter for
the last time. The temperature fluctuations are produced both by
scattering between photons and electrons, and by the gravitational
redshift or blue-shift caused by the evolution of the gravitational
potential between the last-scattering epoch and the present ~\cite{sac}.
On the other hand, the polarization is affected by the
gravitational field only through gravitational lensing
~\cite{Zaldarriaga}, which causes some smearing of the power
spectrum, and also some mixing of the ``electric'' and ``magnetic''
parity modes of the polarization.
In Benson et al. (2001; hereafter Paper I), we
presented a detailed calculation of inhomogeneous reionization of the
intergalactic medium (IGM), using a semi-analytic model of galaxy
formation to predict the ionizing luminosities of stars in galaxies at
different redshifts, and coupling this with an N-body simulation to
predict the spatial distribution of the ionizing sources and of the
ionized gas. In this paper, we use the same model to predict the
effects of reionization on the polarization of the CMB.

Reionization produces interesting effects on CMB temperature and
polarization anisotropies at both first and second order in the
perturbations. At first order, the effects of reionization are the
same as for an IGM with spatially uniform density and ionized
fraction. Density fluctuations in the free electrons around the
reionization epoch produce CMB anisotropies only at second order, as
in the Vishniac effect~\cite{Vishniac}. These second-order
anisotropies are small in amplitude, but nonetheless dominate over the
first-order anisotropies on small angular scales. They are thus
cosmologically interesting as a probe of structure present at
reionization. 
Historically, work on the second-order effects began by considering
the temperature anisotropies in the case of a homogeneous reionization
of the IGM, with the fluctuations in the free electron density being
assumed to follow the variations in the total matter density, the
so-called density modulation model ~\cite{Vishniac,JK}.  However, in
any realistic model, the reionization is expected to be patchy or
inhomogeneous, with some regions already being fully ionized while
others are still neutral, and the ionized regions growing until they
encompass the whole IGM.  Shapiro \& Giroux (1987) and Sato (1994)
investigated the evolution of spherical ionization fronts in a
homogeneous IGM using analytic methods. More recently, numerical
simulations of radiative hydrodynamics have also shown how 
reionization proceeds in an inhomogeneous way \cite{Gnedin2000}.
Gruzinov \& Hu (1998) investigated second-order temperature
anisotropies resulting from patchy reionization using a simple
analytical model in which the universe is reionized in {\em
uncorrelated} spherical patches. More realistic calculations which
relax the assumption of uncorrelated spherical ionized regions have
since been carried out by various authors, using either full
simulations \cite{gnedin} or semi-analytic methods (Paper I,
Springel, White, \& Hernquist 2000, Bruscoli et al. 2000, Valageas,
Balbi, \& Silk 2000). So far, these works have concluded that
second-order effects make significant contributions to the temperature
anisotropy on small scales ($l \gsim 3000$), with peak amplitude in
the range $0.1 \mu$K to $1 \mu$K.

While the second-order temperature anisotropies are well studied,
those of the polarization have received relatively little attention.
Seshadri \& Subramanian (1998) and Weller (1999) discussed this effect
in density modulation and patchy reionization models, respectively.
Here, we consider second-order polarization anisotropies in a more
realistic model than has previously been possible. In our model,
reionization results from photoionization by stars in galaxies, and
the spatial fluctuations in the free electron density are the combined
effect of fluctuations in the total density and fluctuations in the
ionized fraction, thus combining the ``density modulation'' and
``patchy reionization'' effects. We calculate these effects in the
same way as in our previous work (Paper I), by combining a
semi-analytic model of galaxy formation with an N-body simulation of
the distribution of dark matter in the universe to determine the
distribution of ionized regions.  In this paper, we consider only the
scalar mode for the primordial fluctuations.  In the case of
first-order perturbation theory, the scalar mode fluctuations produce
only the electric (E-) mode of polarization \cite{ZS}.  For the
second-order perturbations, however, this may not be the case.  Since
the polarization anisotropies are produced by the coupling between the
large-scale primordial CMB temperature quadrupole anisotropies and the
small-scale density fluctuations in the ionized medium, axisymmetry is
broken and the magnetic (B-) mode is also produced~\cite{Hu}. In
addition to the primordial CMB temperature quadrupole anisotropy,
there is also a kinetic temperature quadrupole in the rest-frame of
the scatterers, produced by the quadratic Doppler effect from the
peculiar velocity of the ionized medium \cite{SunZel}. Coupling
between this kinetic quadrupole and the electron density fluctuations
can likewise produce a second-order contribution to the polarization,
but this contribution is much smaller than the previous one
\cite{Hu}, so we neglect it in the present paper.

This paper is organized as follows. In \S 2, we present the Boltzmann
equation for the first- and second-order polarization anisotropies.
We then derive the E- and B-mode power
spectra by use of the Boltzmann equation.  In \S 3, we give a brief
overview of our model for inhomogeneous reionization and show our
numerical results for both the first- and second-order polarization
power spectra.  \S 4 
is devoted to our conclusions.
Throughout this paper we work in units where $c=1$.

\section{SECOND ORDER EFFECTS FROM THE REIONIZATION EPOCH}
\subsection{Boltzmann Equation and Second Order Polarization}

The evolution of the temperature perturbation, $\Delta_T({\bf x}, {\bf \hat{n}},
\tau)$, and the polarization perturbation, $\Delta_P({\bf x}, {\bf \hat{n}},
\tau)$, are governed by the Boltzmann equation \cite{BEa, PY, MB}.  Here
$\bf x$ is the comoving coordinate,  
$\tau \equiv \int dt / a $ the conformal time, where $a$ is the
expansion factor normalized to unity today, and ${\bf \hat{n}}$ 
the direction of photon propagation.
Following Zaldarriaga \& Seljak (1997), to calculate the polarization
perturbation $\Delta_P$ 
we work in terms of the perturbed Stokes parameters 
$\Delta_{Q\pm iU}$ (see \S 2.2).
In this paper, we do not include tensor perturbations in the metric,
and we derive the 
Boltzmann equation in the conformal Newtonian gauge.  Readers
interested in the synchronous gauge or the transformation between
these two gauges are referred to Ma \& Bertschinger (1995).  In the
Newtonian gauge, the perturbations are specified by two scalar
potentials, $\phi$ and $\psi$, which play the role of the
gravitational potential and the fractional perturbation to the spatial
curvature, respectively. The Boltzmann equation then states that
\begin{equation}
\dot{\Delta}_T+ {\bf \hat{n}}_i \partial_i \Delta_T=\dot{\phi}-
{\bf \hat{n}}_i \partial_i \psi+n_e\sigma_T a(\tau)
\left(-\Delta_T+\Delta_{T0}+{\bf \hat{n}}_i v_{{\rm b} i}-\frac{\sqrt{5\pi}}
{5}\sum_{m=-2}^{m=2}Y_2^m({\bf \hat n})\Pi^{(m)} \right),
\label{realtf}
\end{equation}
\begin{equation}
\dot{\Delta}_{Q\pm iU}+ {\bf \hat{n}}_i \partial_i \Delta_{Q\pm iU}=
n_{\rm e}\sigma_{\rm T}a(\tau)
\left(-\Delta_{Q\pm iU}+\sqrt{\frac{6\pi}{5}}\sum_{m=-2}^{m=2}\ _{\pm2}
Y_2^m({\bf \hat n}) \Pi^{(m)} \right),
\label{realpf}
\end{equation}
where $\ _sY_l^m$ is a spherical harmonic with spin-weight $s$,
and $\Pi^{(m)}$ is defined in terms of the quadrupole components of 
the temperature and polarization perturbations as
\begin{equation}
\Pi^{(m)}({\bf x},\tau) \equiv \Delta_{T2}^{(m)}({\bf x},\tau)+
12\sqrt{6}\Delta_{+,2}^{(m)}({\bf x},\tau)+12\sqrt{6}\Delta_{-,2}^
{(m)}({\bf x},\tau).
\end{equation} 
Here the derivatives are taken with respect to the conformal time,
$v_{\rm b}$ is the velocity of baryons, $n_{\rm e}$ is the free
electron number density, and $\sigma_{\rm T}$ is the Thomson cross
section.
We have also expanded the perturbations in the radiation field 
in spherical harmonics $_sY_m^m$ with 
appropriate spin weight $s$ (see also \S 2.2)
\begin{eqnarray}
\Delta_{T}({\bf x},{\bf \hat{n}},\tau)=\sum_{lm} (-i)^l\sqrt{4\pi(2l+1)}
\Delta_{Tl}^{(m)}({\bf x},\tau)\ Y_l^m({{\bf \hat n}}), \nonumber \\
\Delta_{Q\pm iU}({\bf x},{\bf \hat{n}},\tau)=\sum_{lm} (-i)^l\sqrt{4\pi(2l+1)}
\Delta_{\pm,l}^{(m)}({\bf x},\tau)\ _{\pm2}Y_l^m({\bf \hat n}).
\label{Legendre1}
\end{eqnarray}
When we calculate the perturbations for a mode with wavenumber 
$\bf k$, we define the $_sY_l^m$'s in a coordinate system for which the
$z$-axis is parallel to $\bf k$.
To calculate the first order effect for scalar modes, we can set
$m=0$, as in Zaldarriaga 
\& Seljak (1997), due to the  axisymmetry of the radiation field
around the mode axis for this case. However, their expansion is not
valid for studying the 
second order effect, in which the rotational symmetry around the
wavevector is broken by coupling to other modes.
It is very important to take into account $m\neq 0$ modes for 
the calculating the second order effect,  
otherwise the (artificially imposed) axisymmetry 
guarantees that no magnetic mode can be generated.
This is why Weller (1999), who
assumed axial symmetry following Seshadri \&  Subramanian (1998),
obtained  only the electric mode of polarization for the
second order effect. 
As was found by Hu (2000), who employed the weak-coupling approximation, 
the magnetic mode of polarization is also generated in the case of the
second order effect.
Thus, the expansion in equation (\ref{Legendre1}) is more general and
useful for our calculation. 

On small scales, as pointed out by Kaiser (1984), the contribution to
secondary temperature anisotropies from homogeneous reionization 
tends to cancel 
along the line of sight.  Thus, the first order effect on 
the temperature and polarization anisotropies from the reionization
epoch is suppressed at small angles.  Hereafter, we concentrate on
polarization and develop an equation for the visibility-modulated
effect which is the dominant source on small scales.

Firstly, we write the inhomogeneous distribution of the free electron
number density as
\begin{equation} 
n_{\rm e}({\bf x}, \tau)=\bar{n}_{\rm e}(\tau)[1+\delta_{\rm e}({\bf x}, \tau)],
\end{equation}
where $\delta_{\rm e}$ is the fluctuation in the electron number density
and\ $\bar{}$ denotes the background quantity. Then, equation
(\ref{realpf}) can be rewritten in terms of Fourier modes as follows,
\begin{eqnarray}
\dot{\Delta}_{Q\pm iU}({\bf k},\tau)+ ik\mu \Delta_{Q\pm iU}
({\bf k},\tau)&=&
\bar{n}_e\sigma_T a(\tau)\Biggl(-\Delta_{Q\pm iU}({\bf k},\tau)
-R_{\pm}({\bf k},\tau)  \nonumber \\
& + &  \left .\sum_{m}\sqrt{\frac{6\pi}{5}} 
\ _{\pm 2} Y_2^m({\bf \hat n})
\left(\Pi^{(m)}({\bf k},\tau) +S^{(m)}({\bf k},\tau)\right) \right),
\label{fps}
\end{eqnarray} 
where $S^{(m)}({\bf k},\tau)$ and $R_{\pm}({\bf k},\tau)$ are the 
Fourier modes of the coupling of $\delta_{\rm e}({\bf x},\tau)$ to 
$\Pi^{(m)}({\bf x},\tau)$ 
and $\Delta_{Q \pm iU}({\bf x},\tau)$, respectively. In other words,
\begin{equation}
S^{m}({\bf k},\tau)=
\int \d^3 {\bf p}\delta_{\rm e}({\bf k-p},\tau)\Pi^{(m)}({\bf p},\tau),
\label{defs}
\end{equation}
\begin{equation}
R_{\pm}({\bf k},\tau)=
\int \d^3 {\bf p}\delta_{\rm e}({\bf k-p},\tau)\Delta_{Q\pm iU}
({\bf p},\tau).
\end{equation}
The polarization perturbations at the present time can be obtained by
integrating the Boltzmann equation~(\ref{fps}) along the line of sight
~\cite{ZS}
\begin{eqnarray}
\Delta_{Q\pm iU}({\bf k}, \mu, \tau_0)&=&\int^{\tau_0}_0
\d\tau {\rm e}^{-i k(\tau_0-\tau)\mu} g(\tau)\nonumber \\
&\times& \left (\sqrt{\frac{6\pi}
{5}}\sum_m\ _{\pm 2}Y_2^m({\bf \hat n})\left 
[\Pi^{(m)}({\bf k},\tau)+S^{(m)}({\bf k},\tau)\right ]-
R_{\pm}({\bf k},\tau)\right ),
\label{sol}
\end{eqnarray}
where $\mu={\bf k}\cdot {\bf \hat{n}}$, and $g(\tau)$ is the
visibility function 
defined as 
\begin{equation}
g(\tau)\equiv -\frac{\d\kappa}{\d\tau}{\rm e}^{\kappa(\tau_0)-\kappa(\tau)},
\end{equation}
with $\kappa(\tau) \equiv \int_\tau^{\tau_0}d\tau' a n_{\rm e}
\sigma_{\rm T}$ being the electron-scattering optical depth. The
visibility function has a simple physical meaning, being the
probability that a photon had its last scattering at epoch $\tau$ and
reaches the observer at the present time,
$\tau_0$. Equation~(\ref{sol}) can then be given a simple
interpretation, since $\Pi^{(m)}({\bf k}, \tau)$ acts as a first order
source term while $R_{\pm}({\bf k})$ and $S^{(m)}({\bf k})$ are the
contributions from the second order effect.  To simplify the
calculation, we neglect $R_{\pm}({\bf k})$ and $\Delta^{(m)}_{\pm,2}$
in $S^{(m)}({\bf k})$ because the temperature perturbations typically
dominate over the polarization perturbations.  Furthermore, we assume
the first order temperature quadrupole $\Delta^{(m)}_{T2}$
is {\em uncorrelated}
with $\delta_{\rm e}$ due to the fact that the dominant contributions
to these come from large and
small scales, respectively. That is, we regard the source term for
the second order polarization as
\begin{eqnarray}
S^{(m)}({\bf k},\tau)&\simeq&
\delta_{\rm e}({\bf k},\tau)\int \d^3 {\bf p}\Delta_{T2}^{(m)}
({\bf p},\tau)\nonumber \\
&\equiv& \delta_{\rm e}({\bf k},\tau)Q^{(m)}_2(\tau),
\label{defS}
\end{eqnarray}
with $Q^{(m)}_2(\tau)$ defined as $Q^{(m)}_2(\tau)\equiv \int \d^3 {\bf
p}\Delta^{(m)}_{T2}({\bf p},\tau)$ for convenience. Recall that the scalar
mode in 
linear theory only generates the $m=0$ component in the ${\bf p}$-basis,
i.e. $\Delta^{(0)}_{T2}Y_2^0(\beta,\alpha)$, where $\beta$ and $\alpha$
are the polar and azimuthal angles defining ${\bf \hat n}$ in this
basis. Using the addition theorem, we can project the component in the
${\bf p}$-basis onto the ${\bf k}$- basis (see Ng \& Liu 1999),
\begin{equation}
\sum_{m}Y^{m*}_l({\bf \hat{n}})Y_l^m({\bf
\hat{p}})=\sqrt{\frac{2l+1}{4\pi}}Y_l^0 
(\beta,\alpha),
\end{equation}
 It then follows that
\begin{equation}
Q^{(m)}=\sqrt{\frac{4\pi}{5}}\int\d^3 {\bf p} \Delta^{(0)}_{T2}
({\bf p})Y_2^{m*}(\hat{\bf p}).
\label{Qm}
\end{equation}
Finally, the solution for
$\Delta_{Q\pm iU}$ becomes
\begin{equation}
\Delta_{Q\pm iU}({\bf k}, {\bf \hat{n}}, \tau_0)=\sqrt{\frac{6\pi}{5}}
\int^{\tau_0}_0 \d\tau e^{i k(\tau_0-\tau)\mu}g(\tau)\sum_m
\ _{\pm 2}Y_2^m({\bf \hat n}) X^{(m)}({\bf k},\tau),
\label{sol2}
\end{equation}
where $X^{(m)}({\bf k},\tau)$ equals $\Pi^{(0)}({\bf k},\tau)$ or 
$S^{(m)}({\bf k},\tau)$
for the first and second order contributions respectively.

\subsection{Stokes Parameters and the Power Spectrum}

We follow the approach of Zaldarriaga \&  Seljak (1997) of calculating
the polarization in 
terms of the electric and magnetic modes.
We start with the  description of the polarization perturbation in terms
of the Stokes parameters $Q$ and $U$.  If we consider a wave traveling
in the $\hat{z}$ direction, $Q$ is the difference in intensity
polarized in the $\hat{y}$ and $\hat{x}$ directions, while $U$ is the
difference in the $(\hat{x}+\hat{y})/\sqrt{2}$ and
$(\hat{x}-\hat{y})/\sqrt{2}$ directions.  The circular polarization
parameter, $V$, cannot be produced by scattering, so we will not
mention it further.  Polarization is more complicated than temperature
because the values for $Q$ and $U$ depend on the choice of coordinate
system. Under a right-handed
rotation through an angle $\phi$ around the $\hat{z}$ axis the
temperature is invariant, while 
$Q$ and $U$ transform according to
\begin{equation}
(Q\pm iU)'({\bf \hat{n}})=e^{\mp 2i\phi}(Q\pm iU)({\bf \hat{n}}).
\end{equation}
Thus the functions $Q\pm iU$ have spin $\pm 2$, and should be expanded
in spherical harmonics of spin $\pm2$.


It is possible to produce a rotationally invariant measure of the
polarization field 
if we introduce the spin raising and lowering operators
$\partial\!\!'$ and $\bar{\partial\!\!'}$~\cite{NP}
\begin{eqnarray}
{\partial\!\!'}\eta&=&-(\sin\theta)^s \left[\frac{\partial}
{\partial\theta}
  +i\csc\theta\frac{\partial}{\partial\phi}\right](\sin\theta)^{-s}
\eta, \nonumber \\
\bar{\partial\!\!'}\eta&=&-(\sin\theta)^{-s}\left[\frac{\partial}
{\partial
  \theta} -i\csc\theta\frac{\partial}{\partial\phi}\right]
(\sin\theta)^s\eta,
\label{spinop}
\end{eqnarray}
with $\eta$ as a spin-$s$ field.
If they act on the spin-$s$ spherical harmonics, we have 
\begin{eqnarray}
{\partial\!\!'}\:_{s}Y_{lm}&=&\left[(l-s)(l+s+1)\right]^{1\over 2} 
                            \:_{s+1}Y_{lm},\nonumber \\
\bar{\partial\!\!'}\:_{s}Y_{lm}&=&-\left[(l+s)(l-s+1)\right]^{1\over 2} 
                            \:_{s-1}Y_{lm}.
\label{spinopY}
\end{eqnarray}
The new bases of the rotationally invariant polarization field, called
the electric mode and magnetic mode, are defined as
\begin{eqnarray}
\Delta_{ E}&\equiv&-\frac{1}{2}\left (\baredth^2 
\Delta_{Q+iU}+ \edth^2 \Delta_{Q-iU}\right), \nonumber \\
\Delta_B&\equiv&-\frac{i}{2}\left (\baredth^2 
\Delta_{Q+iU}- \edth^2 \Delta_{Q-iU}\right).
\end{eqnarray}
The values of $\Delta_{ E}$ and $\Delta_B$ at a
particular direction in 
the sky are independent of the coordinate system used to define them
(unlike $\Delta_{Q}$ and $\Delta_{U}$).
We work in the coordinate system where $\hat{k}\ \|\ \hat{z}$ and define
$Q>0$ ($Q<0$) in the direction $\hat{e}_{\theta}(\hat{e}_{\phi}$).  
For a scalar mode in first-order perturbation theory, the radiation
field is axisymmetric around the wavevector. The polarization is
produced by scattering of the quadrupolar component of the radiation
field, and so has only a $\Delta_{Q}$ component,
but no $\Delta_{U}$ component (which would correspond to a
polarization angle at $45^\circ$ to the $\hat{e}_{\theta}$ -
$\hat{e}_{\phi}$ axes), thus
$\Delta_{Q+iU}=\Delta_{Q-iU}=\Delta_Q$. Furthermore, $\Delta_{Q}$ has
no $\phi$-dependence, so $\baredth^2 =\edth^2$
(eq. \ref{spinop}). As a consequence, $\Delta_{ E}=-\edth^2
\Delta_Q=-\baredth^2\Delta_Q$ and $\Delta_B=0$. 
This result is
important because it shows that no magnetic mode can be produced by
scalar modes (density perturbations) in first-order perturbation theory.
However, as we will show later, the same situation does not occur in the
second-order contribution to the polarization because the coupling
between modes breaks 
axisymmetry. 

Again, following Zaldarriaga \& Seljak (1997), the power spectra
of the electric mode and magnetic mode can be defined as 
\begin{equation}
C_{(E,B)l}\equiv\frac{1}{2l+1}\frac{(l-2)!}{(l+2)!}\sum_m \int \d^3 
{\bf k} <|\Delta_{(E,B)l}^{(m)}({\bf k},\tau=\tau_0)|^{2}>,
\label{CEL}
\end{equation}
where $\Delta^{(m)}_{(E,B)l}$ can be extracted using
\begin{equation}
\Delta^{(m)}_{(E,B)l}({\bf k})= \int \d\Omega 
Y_l^m(\theta,\phi) 
\Delta_{(E,B)}({\bf k}).
\label{del}
\end{equation}

To calculate equation (\ref{del}), we can apply the spin raising and
lowering operators (eq. \ref{spinop} or \ref{spinopY}) to the
polarization perturbations in equation (\ref{sol2}) twice. But the plane
wave $e^{ik(\tau_0-\tau)\mu}$ itself carries an angular dependence, thus
we expand the plane wave in a
series of spherical waves using the Rayleigh equation
\begin{equation}
e^{ik(\tau_0-\tau)\mu}=\sum_l (-i)^l \sqrt{4\pi (2l+1)}j_l[k(\tau_0-
\tau)] Y^0_l({{\bf \hat n}}),
\label{ray}
\end{equation}
where $j_l[k(\tau_0-\tau)]$ is the spherical Bessel function. 
Furthermore, we calculate the product  of two spherical
harmonics with spin by using the Clebsch-Gordan relation~\cite{Saku}
\begin{eqnarray}
\ _{s_1}Y_{l_1}^{m_1}\ _{s_2}Y_{l_2}^{m_2}&=&\frac{\sqrt{(2l_1+1)
(2l_2+1)}}{4\pi}\sum_{lms}<l_1,l_2,m_1,m_2|l,m>\nonumber \\
&& \times <l_1,l_2,-s_1,-s_2|l,-s>
\sqrt{\frac{4\pi}{2l+1}}\ _{s}Y_{l}^{m}.
\end{eqnarray}
The expression for $\Delta_{(E,B)l}^{(m)}({\bf k},\tau)$ then becomes
\begin{equation}
\Delta_{(E,B)l}^{(m)}({\bf k},\tau)=\frac{3}{2}\sqrt{\pi}\frac{(l+2)!}
{(l-2)!}\sqrt{2l+1}\int_0^{\tau_0}\d\tau g(\tau)
X^{(m)}({\bf k},\tau)T_{(E,B)l}^{(m)}(kr),
\label{DEL2}
\end{equation} 
with $r=\tau_0-\tau$.
In Table~1, we list the results of $T_{(E,B)l}^{(m)}(kr)$, in which the
orthonormality relation 
\begin{equation}
\int d\Omega \ _sY_l^{m*}\ _sY_{l'}^{m'}=\delta_{ll'}\delta_{mm'},
\end{equation}
and the recursion relations of spherical Bessel functions
\begin{equation}
\frac{j_l(x)}{x}=\frac{1}{2l+1}[j_{l-1}(x)+j_{l+1}(x)]
\end{equation}
have been used.
Finally, the power spectrum can be obtained by substituting 
equation (\ref{DEL2}) into (\ref{CEL}),
\begin{equation}
C_{(E,B)l}=(4\pi)^2\frac{9}{16}\frac{(l+2)!}{(l-2)!}
\sum_m\int k^2 \d k \left\langle \left |\int \d\tau g(\tau) X^{(m)}({\bf k},\tau) 
T_{(E,B)l}^{(m)}(kr)\right |^2 \right\rangle.
\label{fsol}
\end{equation}

In the case of the second-order effect, eq.(\ref{fsol}) can be
rewritten as an expression involving the power spectrum of the
fluctuations in the free electron density $\langle\delta_e({\bf
k},\tau) \delta^*_e({\bf k},\tau')\rangle$, connecting the two
different times $\tau$ and $\tau'$. When we evaluate eq.(\ref{fsol})
numerically, we neglect the effect of the difference in times in the
power spectrum, and so replace it by $\langle\delta_e({\bf
k},\bar\tau) \delta^*_e({\bf k},\bar\tau)\rangle$, evaluated at a
single time $\bar\tau$ which is an average of $\tau$ and $\tau'$. 
This should
be an adequate approximation if the dominant wavelengths in
eq.(\ref{fsol}) are small compared to the Hubble radius.

\begin{table}[t]
\begin{tabular}{|l|c|c|} \hline
$m$ & $T_{El}^{(m)}$ & $T_{Bl}^{(m)}$ \\ \hline
$0$ & $(-i)^l\frac{j_l(kr)}{(kr)^2}$ & 0 \\ \hline
$\pm 1$ & $\mp (-i)^l\frac{1}{(2l+1)kr}\sqrt{\frac{1}{6l(l+1)}}
[lj_l(kr)-(l+1)j_{l-1}(kr)]$ & $\pm \sqrt{\frac{(-i)^l}{6l(l+1)}}
\frac{j_l(kr)}{(kr)^2}$ \\ \hline
$\pm 2$ & $\pm \frac{(-i)^l}{2l+1}\sqrt{\frac{(l-2)!}{24(l+2)!}}
\left(\left[\frac{(l+2)(l+1)}{2l-1}+\frac{l(l+1)}{2l+3} 
+6\frac{(2l+1)(l-1)
(l+2)}{(2l-1)(2l+3)}\right] \right.
 $ & $\mp \frac{(-i)^l}{2l+1}
\sqrt{\frac{(l-2)!}{6(l+2)!}}\left[(l+2)j_{l-1}(kr) \right.$ \\ 
 &$\left.\times j_l(kr) -(l+2)(l+1)\frac{j_{l-1}
(kr)}{kr}+l(l-1)\frac{j_{l+1}(kr)}{kr}\right )$ & $ \left .
-(l-1)j_{l+1}(kr)
\right]$ \\ \hline 
\end{tabular}
\caption{$T_{(E,B)l}^{(m)}$ in equation ({\ref{DEL2}})}
\end{table}%

\section{NUMERICAL RESULTS}

To calculate the polarization anisotropy spectrum using the results
discussed above, we use the publicly available code CMBFast~\cite
{SZ}. We modify the ionization history in this code to follow the more
realistic case from our previous work (Paper I), in which the
reionization history of the universe is determined by a semi-analytic
model of galaxy formation. The semi-analytic model is that of Cole et
al. (2000), which includes the following processes: formatinon and
merging of dark matter halos through hierarchical clustering;
shock-heating and radiative cooling of gas within these halos;
collapse of cooled gas to form galactic disks; star formation in disks
and feedback from supernova explosions; galaxy mergers; chemical
evolution of the stars and gas; and luminosity evolution of stellar
populations based on stellar evolution codes and model stellar
atmospheres. The model has been shown by Cole et al. to agree well
with a wide range of observed properties of galaxies in the local
universe.  In Paper I we used this model to calculate the ionizing
luminosities of galaxies at different redshifts, including the effects
of absorption by interstellar gas and dust on the fraction of ionizing
photons escaping, and followed the propagation of the ionization
fronts around each galaxy.  To find the ionizing luminosity, we first
calculate the rate at which ionizing photons are being produced by
stars in the galaxy, then apply an attenuation due to dust, and
finally allow a fraction $f_{\rm esc}$ of the remaining photons to
escape into the IGM.  The mass of ionized hydrogen in each spherical
ionization front is found by integrating the equation \cite{SG}
\begin{equation}
{1 \over {\rm m}_{\rm H}} {{\rm d}M \over {\rm d}t} = S(t) - \alpha_{
\rm H}^{(2)} a^{-3} f_{\rm clump} n_{\rm H} {M \over {\rm m}_{\rm H}},
\label{ion_cgs}
\end{equation}
where $n_{\rm H}$ is the comoving mean number density of hydrogen
atoms (total, \HI\ and \HII) in the IGM, $m_{\rm H}$ is the mass of a
hydrogen atom, $a(t)$ is the scale factor of the universe normalized
to unity at $z=0$, $t$ is time (related to the conformal time by
$dt=d\tau/a$), $S(t)$ is the rate at which ionizing photons are being
emitted and $\alpha_{\rm H}^{(2)}$ is the case B recombination
coefficient. The clumping factor $f_{\rm clump} \equiv \langle
\rho_{\rm IGM}^2\rangle/\bar{\rho}_ {\rm IGM}^2$ gives the effect of
clumping on the recombination rate of hydrogen in the IGM. A larger
$f_{\rm clump}$ increases the recombination rate resulting in a delay
of the reionization epoch. We use the clumping factor $f_{\rm
clump}^{\rm (halos)}$ as defined in Paper~I. By summing over the
ionized volumes due to all galaxies, we can calculate the fraction of
the IGM which has been reionized at any redshift. Having this
reionization history, we can then obtain the first order power
spectrum of the polarization anisotropies without making any
assumptions about the spatial distribution of the ionized gas.  We
consider only the scalar mode of the primordial 
perturbations, for which the radiation field for each ${\bf k}$-mode is axially
symmetric. Thus, we set $m=0$ in equation (\ref{fsol}) and find no
magnetic mode can be produced (see Table.~1).

The second order effect is more complicated, as the source term
contains $\delta_{\rm e}$ and $\Delta^{(m)}_{T2}$, i.e., the $m=0, \pm
1$ and $\pm 2$ components of the temperature quadrupole must be
considered.  The time evolution of the temperature quadrupole
components with different $m$ in the ${\bf k}$-basis is obtained using
equation (\ref{Qm}), and $\Delta_{T2}^{(0)}$ is calculated using the
modified CMBFast code. This leaves only the power spectrum of the
electron density fluctuations $\delta_{\rm e}$ unknown, which should
be calculated from the distribution of ionized regions with different
sizes and shapes and with a correlated spatial distribution.  An exact
calculation would require a high resolution numerical simulation with
gas dynamics and radiative transfer included, and so would require
very large amounts of computing time.  Here we instead calculate the
density field $\delta_{\rm e}$ of the ionized gas using the simpler
approach of Paper~I, in which the semi-analytic galaxy formation
model is combined with a high-resolution N-body simulation of the dark
matter.  The simulation volume, which is a box of comoving length $141.3 h^{-1}
{\rm Mpc}$ and contains $256^3$ dark matter particles each of mass
$1.4 \times 10^{10} h^{-1} M_{\odot}$, is divided into $256^3$ cubical
cells of comoving size $0.55 h^{-1}
{\rm Mpc}$. Each dark matter halo in the simulation containing more than 10
particles is populated with galaxies according to the semi-analytic
model, and the sum of their ionizing luminosities placed in a source
at the center of mass of the halo. Ionizing photons which originate
from galaxies in lower mass halos (which are not resolved in the
simulation) are added in by assuming that these unresolved halos trace
the remaining mass of the simulation (i.e. that mass which is not part
of a resolved halo). In Paper I we demonstrated that the locations of
unresolved sources do not significantly affect the resulting
anisotropy spectrum. We determine which regions of the simulation box
become ionized by using one of the five toy models A-E listed below,
which span the likely range of possible behaviour. In all cases, the
total mass of ionized hydrogen in the simulation box is the same, and
is forced to equal that for a large-scale homogeneous
distribution with the specified IGM clumping factor $f_{\rm clump}$,
calculated by use of equation (\ref{ion_cgs}). We assume that the
total gas density in the IGM traces the dark matter, thus neglecting
the effects of pressure in the IGM. Hu (2000) has shown that this is a
good approximation for anisotropies with $l\lsim 10^4$.

{\bf Model A (Growing front model)} Ionize a spherical volume around
each halo with a radius equal to the ionization front radius for that
halo calculated assuming a large-scale uniform IGM. Since in the
simulation the IGM is \emph{not} uniform, but is assumed to trace the
dark matter, and also because some spheres will overlap, the ionized
volume calculated in this way will not contain the correct total
ionized mass. We therefore scale the radius of each sphere by a
constant factor, $f$, and repeat the procedure. This process is
repeated, with a new value of $f$ each time, until the correct total
mass of hydrogen has been ionized.

{\bf Model B (High density model)} In this model, we ignore the
positions of halos in the simulation. Instead, we simply rank the cells
in the simulation volume by their density. We then completely ionize
the gas in the densest cell. If this has not ionized enough \HI\ then we
ionize the second densest cell. This process is repeated until the
correct total mass of \HI\ has been ionized.

{\bf Model C (Low density model)} This is like model B, except that we
begin by ionizing the least dense cell, and work our way up to cells
of greater and greater density. This model mimics that of
Miralda-Escud\'e, Haehnelt \& Rees (2000).

{\bf Model D (Random spheres model)} As Model A, except that the
ionized spheres are placed at completely random positions in the
simulation volume, rather than on the dark matter halos to which they
belong. By comparing to Model A this model allows us to estimate the
importance of the spatial clustering of dark matter halos.

{\bf Model E (Boundary model)} Ionize a spherical region around each
halo with a radius equal to the ionization front radius for that
halo. This may ionize too much or not enough \HI\  depending on the
density of gas around each source. We therefore begin adding or
removing cells at random from the boundaries of the already ionized
regions until the required mass of \HI\ is ionized.

From the results of the simulation, we calculate $\delta_{\rm e}$ as
\begin{equation}
\delta_{\rm e}({\bf x},\tau)=\frac{x_e({\bf x},\tau)(1+\delta)}
{\bar{x}_e(\tau)}-1,
\end{equation}
where $x_{\rm e}$ is the ionized fraction (which we take to equal 1 in
ionized regions and 0 in neutral regions) and $\delta$ is the dark
matter overdensity (i.e. we assume that fluctuations in the gas
density follow those in the dark matter). Here $\bar{x}_e$ is defined
as the mass-averaged ionized fraction in the IGM.

We fix the cosmological parameters to be $\Omega_0=0.3$,
$\Lambda_0=0.7$, Hubble constant ${\rm H}_0=70$ km/s/Mpc and
$\sigma_8=0.9$.  We will consider the effects of varying $f_{\rm esc}$
and $\Omega_{\rm b}$ on the polarization anisotropies and also examine
the polarization power spectrum in all five toy models for the
distribution of ionized regions.  In Fig.~1 we plot the visibility functions of
the ionization histories in our simulation for different $f_{esc}$
(panel (a)) and different $\Omega_{\rm b}$ (panel (b)).  We find that
the visibility function depends strongly on $f_{\rm esc}$, and also on
$\Omega_{\rm b}$. The value of $\Omega_{\rm
b}$ determines the cooling rate (and so star formation rate) in our
model of galaxy formation and also alters the recombination rate in
the IGM, and so affects the time at which reionization occurs (panel
(b)).
 Note that when we vary $\Omega_{\rm b}$ and $f_{\rm
esc}$, we also vary the fraction of brown dwarfs in the IMF used in
the galaxy formation model, in such a way that the model always fits the
``knee'' of the observed $H\alpha$ luminosity function of galaxies at
$z=0$ (see Paper~I for more details). Therefore the production of
ionizing photons does not simply scale with $\Omega_{\rm b}$.

%
In Fig.~2, we show how the first-order polarization anisotropies are
affected by scattering by free electrons at the reionization epoch,
calculated using the
modified CMBFast code.  We find, first, that rescattering at the
reionization epoch generates a new anisotropy at {\em large scales}
because the horizon has grown to a much larger size by reionization
than it had at recombination ($z\simeq 1100$).  More specifically, the
location of the new peak reveals the horizon size at last scattering,
and its height reveals the duration of last scattering, i.e., this new
peak is sensitive to the optical depth produced by reionization.
To see how the distortion of the primordial polarization depends on
the optical depth $\kappa_i$ back to the start of reionization, we
plot the resulting power spectrum for various values of $f_{\rm esc}$
in panel (a).  We find that the boost in the large-scale power depends
strongly on the value of $f_{\rm esc}$. (The optical depth is $0.034$,
$0.017$ and $0.014$ for $f_{\rm esc}=1.0, 0.1$
and $0.05$ respectively, for $\Omega_{\rm b}=0.02$.)  The location of
the new peak depends on the reionization epoch, which is
characterized by the redshift $z_i$ of the peak of the visibility function
as shown in Fig.~1.  From numerical simulations of the Boltzmann code
for different reionization epochs and cosmological parameters, we
obtain a fitting formula for the peak location $l_{\rm peak}$ as a
function of the reionization epoch $z_i$:
\begin{equation}
l_{\rm peak} = 0.74 (1 + z_i)^{0.73}\Omega_0^{0.11}. 
\end{equation}
This fit is consistent with the peak locations shown in 
Fig.~2. 
On the other hand, the first-order polarization fluctuations are suppressed on
small scales by rescattering since a fraction of the photons coming
from the recombination epoch is scattered after reionization. The tiny
optical depth for rescattering causes little erasure of power on
small scales, but the new peak reaches an amplitude of
$\sim$0.05--0.1$\mu$K. From Fig.~2(b), we can see that the suppression
also depends on $\kappa_i$.

Let us discuss now the second-order effect. 
In Fig.~3, we show the contribution to $C_{El}$ and $C_{Bl}$ from
{\em each} value of $m$ for Model A. 
For $m=0$, only the E-mode is generated
because $T^{(0)}_{Bl}=0$ (see Table.~1). Our numerical results show 
that the E-mode (the B-mode) from $m=\pm 2$ ($m=\pm 1$) dominates 
the one from  $\pm 1$ ($\pm 2$) by more than four orders of magnitude.
However, the total power spectra of the E- and B-modes are almost 
exactly the same.
The difference is less than $10^{-6}$, which may be caused by
numerical errors.  These results are consistent with previous work by
Hu (2000), in which the weak-coupling approximation was employed. (The
weak-coupling approximation involves making an analytical
approximation for the line-of-sight integrals, while in
eq.(\ref{fsol}) we instead calculate the integrals over $\tau$ and $k$
numerically.) The reason for the equality of the total E- and B-mode
power spectra is essentially that the first-order quadrupole whose
scattering produces the polarization is dominated by large scales, and
so has a random orientation relative to the small-scale fluctuations
in the electron density (c.f eqn.(\ref{defS})). Scattering of the
quadrupole by the small-scale fluctuations therefore on average
excites E- and B- modes equally.

Numerically, the amplitude of the second-order signal is found to be
about $10^{-8}$K. This can also be understood as follows: from
(eq.~\ref{sol2}), the amplitude of the second order polarization
anisotropies is roughly $\Delta_{Q\pm iU}\sim \int d\tau g(\tau)
Q_2^{(m)} \delta_e \sim \kappa Q_2^{(m)} \delta_e$. In our
calculation $\kappa, Q_2^{(m)}$ and $\delta_e$ are typically
on the order of $10^{-2}, 10^{-5}{\rm K}$ and $10^{-1}$,
respectively. The order-of-magnitude of the second-order contribution
then follows.

The relation between the angular wavenumber $l$ and the comoving
wavenumber $k$ of the density fluctuations at a given redshift is
\begin{equation}
l\sim k r(z) = k\int_0^z\frac{dz}{H_0\sqrt{\Omega_0(1+z)^3+1-\Omega_0}}.
\end{equation}
where $r(z)$ is the standard radial coordinate distance in the
Robertson-Walker metric, and the final expression assumes a flat
cosmology. For our standard choice $\Omega_0=0.3$, $\Lambda_0=0.7$, we
find $r(z)\approx 6000 h^{-1} {\rm Mpc}$ for reionization at $z_i \sim
5-10$. Thus, electron density fluctuations at reionization on comoving scale
$\Delta x$ produce second-order anisotropies at $l \sim 6000\, (h^{-1}
{\rm Mpc}/\Delta x)$. 


In Fig.~4, we plot the second-order power spectrum of the polarization
in the five toy models with fixed extreme escape fraction $f_{\rm
esc}=1$ and $\Omega_b=0.02$.  Although the shapes of the curves are
all very similar, their amplitudes are different. Note that the
reduction in power above $ l \sim 10,000$ is artificial and due to the
limited resolution of the N-body simulation we use (the density field
of the ionized gas is calculated on a grid with cell size $0.55 h^{-1}
{\rm Mpc}$, corresponding to $ l \sim 10^4$). On the other hand, the
finite size of the simulation box ($256 h^{-1} {\rm Mpc}$) affects the
power spectrum for $l$ below a few hundred.
We see that the amplitude  of the power spectrum around the peak ($l\approx
10,000$) varies by a factor $\approx 2.5$, depending on which of the
models A--E is used.
In Paper I we argued that the amplitude of the
curves is affected by the strength of the correlations of
$\delta_e$ present in each model.  As a result, the ``high density''
model (B) is the most strongly 
correlated and has the highest amplitude, and conversely the ``low
density'' model (C) has the lowest amplitude. 


In Fig.~4, we also compare our results to the analytical toy model of
Gruzinov \& Hu (1998), in which the reionization is described by three
free parameters. In their model, each luminous source is assumed to
ionize a spherical region with fixed comoving radius $R$, the first
source appears at redshift $z_i$, and new sources turn on at a
constant rate until reionization is complete after an interval $\delta
z$.  An artificial assumption is made that luminous sources appear at random
locations in space, so there are no correlations between the
positions of the ionized spheres.  Assuming that the spheres remain
ionized forever, the fractional ionization increases with increasing
number density of ionized spheres during $\delta z$ until the universe
is completely ionized. In Paper~I, we chose $R=0.85h^{-1}$ Mpc,
$z_i$=11 and $\delta z=5$ in the Gruzinov \& Hu model to match the
peak in the power spectrum of secondary {\it temperature} anisotropies
predicted for our model~E. We choose the same parameters here and find
that the toy model likewise matches the peak in the spectrum of {\em
polarization} anisotropies for model~E.  For small $l$, little power
is generated in the Gruzinov \& Hu model because, by design, the
patches are uncorrelated.

To further clarify what determines the {\em shape} of the second-order
anisotropy spectrum in our models, we carried out the following
additional tests, shown in Fig.~5. The first test was to force the
ionized fraction $x_e$ to be uniform and equal to the same mean value
as before, so that fluctuations $\delta_e$ in the free electron
density are then simply equal to fluctuations $\delta$ in the {\em
total} density. In this case, the angular power spectrum has an almost
identical shape to models~A-E, with amplitude about equal to that for
models C, D or E, but lower than that for models A or B (by factors of
about 1.7 and 2.5 respectively). In the next two tests (labelled
``random'' in Fig.~5), the {\em total} gas density was forced to be
uniform (i.e. we set $\delta=0$), and put bubbles down at random
positions, so that fluctuations in $\delta_e$ resulted only from the
patchiness of the reionization. In one case, the bubble radii were
chosen from the size distribution predicted by our galaxy formation
model (this case is thus similar to model~D, except that model D had
$\delta\neq 0$). In the other case, all bubbles were given the same
comoving radius of $0.62 h^{-1}{\rm Mpc}$, which corresponds to the
mean bubble radius (weighted by bubble volume) predicted by the galaxy
model, at the redshift corresponding to the peak of the visibility
curve. Both of these cases give power spectra with shapes (at large
scales) similar to the analytical Gruzinov \& Hu model, and completely
different from when fluctuations in $\delta$ are included. In the
final test (labelled ``clustered'' in Fig.~5), we again forced the
bubble radii to be equal at a given redshift, but placed them on
random halos, and included the fluctuations in the total gas
density. The starting value for the radii in this last case was again
$0.62 h^{-1}{\rm Mpc}$, but the spheres were then grown by a uniform
factor at each redshift to produce the correct mean ionized fraction,
as in model A. The power spectrum in this case is almost the same as
in model~A, showing that the distribution in bubble sizes in the
latter case does not have much effect.


From these tests, we conclude that in our models A-E, the {\em shape}
of the power spectrum on scales large compared to the typical size of
the ionized bubbles is determined primarily by the correlations in
total gas density. However, in the case of {\em patchy} reionization,
the {\em amplitude} depends on the spatial distribution of these
patches, which produces {\em biasing} for the correlations in the
ionized gas density relative to those in the total gas density, which
in turn boosts the amplitude of the polarization fluctuations. For our
models A-E, the boost in amplitude of the power spectrum, relative to
the uniform $x_e$ case, varies from about a factor 1 to 2.5, the
largest boost resulting from the case where the densest cells are
ionized first (model~B).

The amplitude of the second-order polarization power spectrum also
depends on the value of $\Omega_{\rm b}$, which affects the visibility
function in equation~(\ref{DEL2}). In our calculation, $\Omega_{\rm
b}$ is set to be $0.02$, which is lower than the current estimates
based on Big Bang Nucleosynthesis which imply $\Omega_{\rm b}
h^2=0.020 \pm 0.002$~\cite{Burles}.  In Fig.~6, we increase
$\Omega_{\rm b}$ to $0.04$ (which results in $\kappa_i$ increasing to
0.048).  If the evolution of the ionized regions were the same in both
models, we would expect the amplitude to increase a factor of four
(eq.~\ref{sol2}).  However, the delay of reionization resulting from
this increase in $\Omega_{\rm b}$, as was shown in Fig.~1, reduces the
factor to $2.8$.

So far we have set $f_{\rm esc}=1.0$ only. Whilst this may in fact be
a plausible value for high-redshift galaxies based on recent
observations (Steidel Pettini \& Adelberger 2000), it is
instructive to examine the effects of changing $f_{\rm esc}$.  Using
Model A 
we examine the effects on the second order polarization 
anisotropies of varying $f_{\rm esc}$. The results are shown in
Fig.~7.  In contrast to the case of first-order anisotropies, the amplitude of
the second-order anisotropies increases only weakly with increasing $f_{\rm
esc}$, with the extreme value $f_{esc}=1.0$ differing from the others
by only a factor of about $1.6$.  This is because the source term for the
first-order polarization anisotropies is the visibility function
multiplied by the temperature quadrupole, $g(\tau)\Delta_{T2}$. Since
the temperature quadrupole changes slowly with time, we can say that
the power spectrum is mainly determined by the visibility function.
However, the source term for the second order anisotropies is the
visibility function multiplied by $\delta_{\rm e}$, whose time
evolution is roughly proportional to $1/(1+z)$. Thus, the signal in
the second-order anisotropies is more strongly weighted to low $z$, where the
differences between models with different $f_{\rm esc}$ are less
significant. 

We also plot the first-order power spectrum as heavy solid lines
in Figs.~4-7. This shows that the second order signal
dominates over the first order signal on small scales $(l\gsim 6000)$.
This will be useful for a high resolution 
experiment since we will be able to distinguish the secondary signal
from the primordial one.

\section{CONCLUSIONS}

We have investigated the polarization anisotropies created during the
reionization epoch, for a realistic model of reionization by galaxies,
in particular focusing on the second-order effect.  It is found that
the boost in the large-scale polarization anisotropy from the
first-order effect is very sensitive to the optical depth due to
reionization, with larger optical depths giving larger boosts. Using a
semi-analytic model of galaxy formation to calculate reionization
(Paper I), the optical depth from the present to the recombination
epoch, $\kappa_i$, is predicted to be in the range $0.014-0.048$,
for $\Omega_b$ in the range $0.02-0.04$, and the escape fraction of
ionizing photons from galaxies $f_{\rm esc}$ in the range
$0.05-1.0$. This first-order effect causes a new peak in the
polarization at large scales with amplitude around $0.05-0.15
\mu$K. The position of the peak is determined by the size of the
horizon at reionization. If this peak can be detected by future
experiments such as Planck, it would offer valuable evidence to guide
our understanding of reionization of the universe.

We have also studied the second-order polarization effects resulting
from the coupling of fluctuations in the free electron number density,
$\delta_{\rm e}$, with the quadrupole of the temperature anisotropies.
We obtained $\delta_{\rm e}({\bf x},t)$ by combining the semi-analytic
model of galaxy formation with a high resolution N-body simulation, as
in our previous work (Paper I). The semi-analytic model determines the
average ionized fraction at each redshift, but the spatial
distribution of ionized gas is based on using the simulation to
determine the locations of the ionizing sources and density field of
the IGM. We determine which regions of the IGM are ionized using one
of five toy models. We summarize our results for the second-order
polarization anisotropies as follows: (1) The second order effect
dominates over the first-order effect on small scales ($l > 6000$).
(2) The B-mode of polarization is induced as well as the E-mode.  The
angular power spectra of these two modes $C_{El}$ and $C_{Bl}$ are the
same. (3) The shapes of the $C_{(E,B)l}$ are very similar in all of
the toy models we considered for the spatial distribution of the
ionized regions, but differ considerably from the toy model of
Gruzinov \& Hu (1998), having much more power on larger angular
scales.  The reason for this difference is the spatial correlations of
the total gas density in our model, which produce corresponding
large-scale correlations in the density of ionized gas also. The shape
of $C_{(E,B)l}$ on large scales is determined mostly by the
correlations in the total gas density, but its amplitude is sensitive
to the geometry of the ionized regions, which determines the biasing
factor of the correlations in ionized density relative to those of the
total gas density.  (4) The difference in amplitude between our
different toy models for the spatial distribution of ionization is
large in the second order effect, for example, Models B and C differ
by a factor of $2.5$. Thus, they provide a very important constraint
on galaxy formation and reionization, in spite of the fact that the
shapes of all curves are similar.  The cosmological parameters which
also affect the amplitude of the second-order effect will be
determined by forthcoming precise measurements of CMB anisotropies by
MAP and PLANCK. Therefore we expect that when the second order
polarization is observed, its amplitude will provide an important
signature of the galaxy formation and reionization processes.
(5) The amplitude also depends on
the value of $\Omega_{\rm b}$.  If the evolution of the ionized
regions were the same for any value of $\Omega_{\rm b}$, we would
expect the amplitude to be proportional to $\Omega_{\rm
b}^2$. However, the scaling in our model is somewhat weaker than this,
because
increasing the value of $\Omega_{\rm b}$ delays reionization. In our
calculation, the second-order power spectrum for an $\Omega_{\rm
b}=0.04$ model has an amplitude 2.8 times greater than for an
$\Omega_{\rm b}=0.02$ model.  (6) The amplitude of $C_{(E,B)l}$ depends
only weakly on $f_{\rm esc}$ (and so on the redshift of
reionization). 
Since the amplitude of the power spectrum for the extreme value of
$f_{\rm esc}=1.0$ differs from that with $f_{\rm esc}=0.1$ only by a
factor of 
about $1.5$, the results in Fig.~4 and Fig.~6 should be only weakly
dependent on the true value of $f_{\rm esc}$.

While second-order effects can make a significant contribution to the
temperature anisotropy ($\sim$0.1--1$\mu$K), the second-order
polarization anisotropies generated by inhomogeneous reionization are,
as expected, very small.  We conclude that the signals are typically
of order 10nK on arcminute scales, comparable to estimates from other
authors studying Vishniac-type polarization~\cite{SS} or toy models
of patchy reionization \cite{Weller}.  A signal of this amplitude is
below the detectability limits of the Planck Surveyor mission, which
is the most accurate experiment in the near future. However, it is
worthwhile to note that this second order effect generates the same
power in the E- and B-modes. Thus a detection of B-mode polarization
on arcminute scales or below is essential to distinguish the
second-order effect from other possible effects.  Since it provides
more information about galaxy formation and the process of
reionization than the first-order effect, detection of this signal
should be a key aim of a post-Planck experiment with increased
sensitivity and resolution in the next decade. Among currently
planned instruments, the most useful one for detecting the
second-order signal is probably the mm-wavelength interferometer ALMA,
which will be able to measure CMB fluctuations on arcminute
scales. Since the second-order signal is very small, polarized
Galactic foregrounds, from dust (e.g. \cite{prunet}.) and synchrotron
emission (e.g. \cite{bacci}), are a concern. However, the amplitude
of these foregrounds on arcminute scales is currently unknown, as
observational estimates have been obtained only for larger scales ($l
\lsim 100-1000$). The detectability of the CMB polarization signal can
thus only be decided by future observational work.

\acknowledgments
We thank U. Seljak and M. Zaldarriaga for the use of CMBFast, the
Virgo Consortium for making available the GIF N-body simulations used
here, and Shaun Cole, Carlos Frenk and Carlton Baugh for allowing us
to use their model of galaxy formation. GCL thanks N. Seto
for useful discussion and acknowledges the fellowship of Interchange
Association. NS, AN and CGL acknowledge kind hospitality of 
Carlos Frenk and the physics department of University of Durham during 
the TMR network meeting.
NS is supported by the Sumitomo fundation. CGL acknowledges support at
SISSA from COFIN funds from MURST and from ASI.
AN and AJB acknowledge the support of the
EC RTN network ``The Physics of the Intergalactic Medium''.

\begin{figure}
\plotone{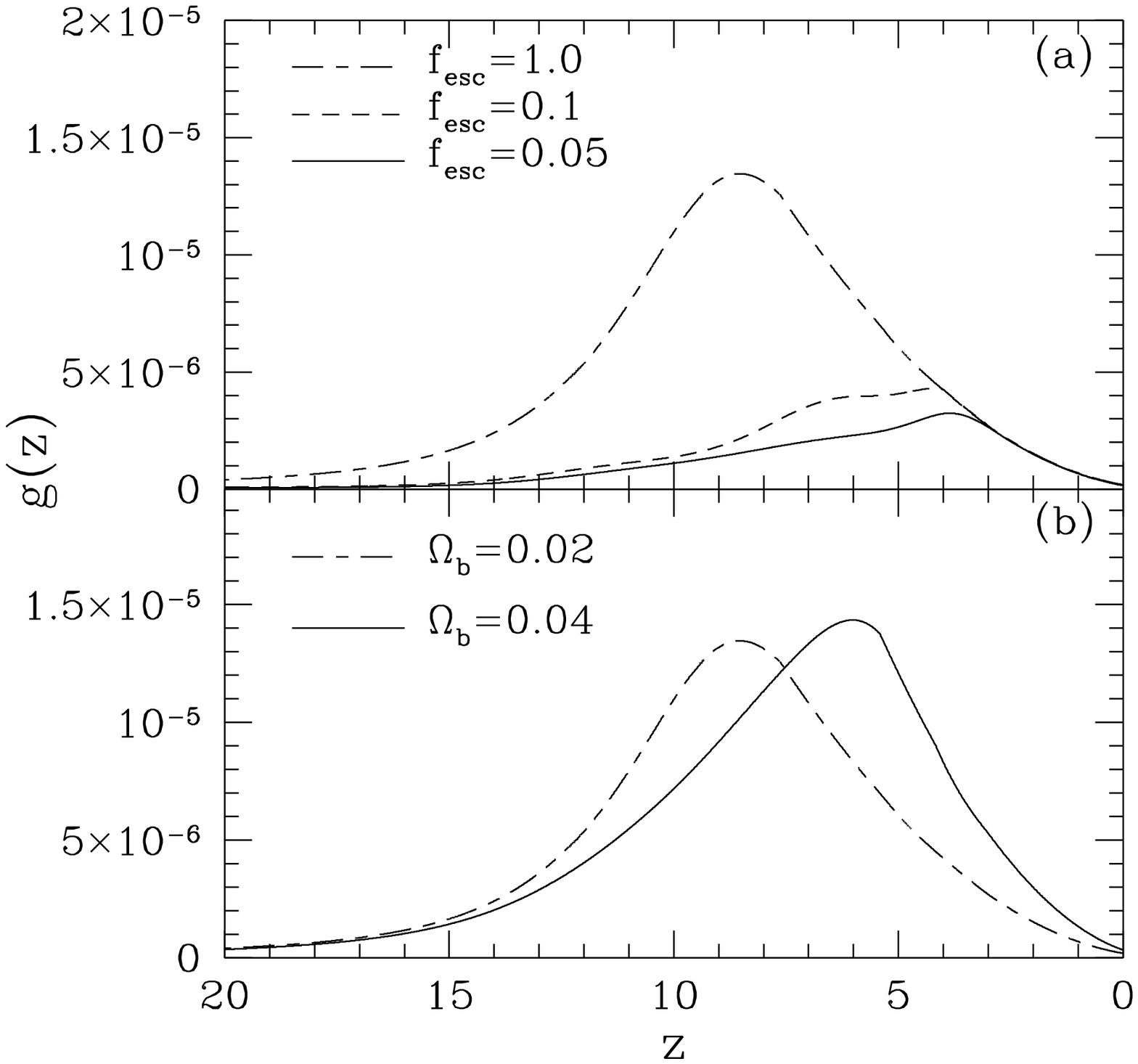}
\caption{Visibility function for various escape fractions $f_{\rm
esc}$ (panel (a)) and baryon fractions $\Omega_{\rm b}$ (panel (b)) in
$\Lambda$CDM models with $\Omega_0=0.3$, $h=0.7$ and $\Lambda=0.7$. In
panel (a), $\Omega_{\rm b}$ is fixed at $0.02$, and in panel (b), $f_{\rm
esc}$ is set to $1.0$. }
\label{Fig. 1}

\end{figure}

\begin{figure}
\plotone{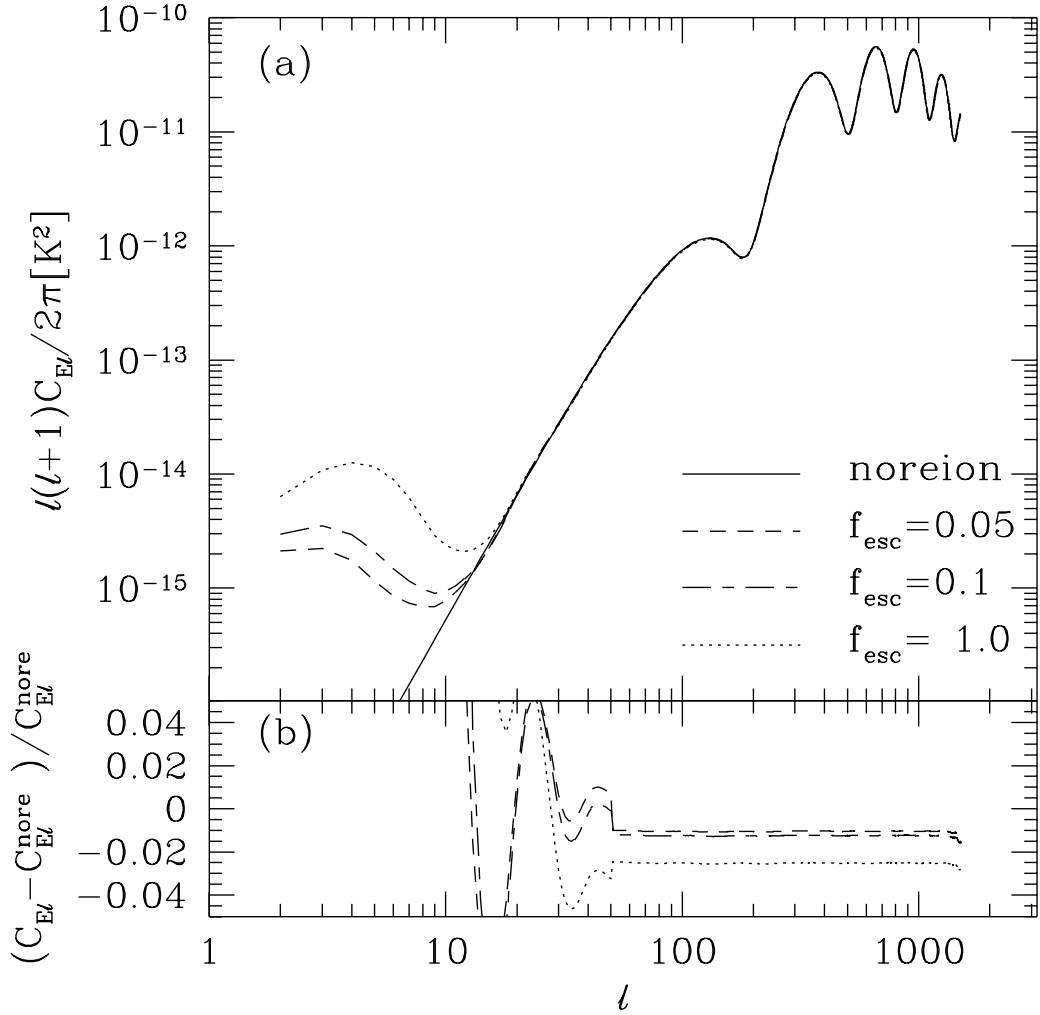}
\caption{
The modification of the first-order
polarization power spectrum by reionization.  (a) The angular power
spectrum of the first-order polarization produced by the reionization
histories with $\Omega_{\rm b}=0.02$ and different escape fractions
$f_{\rm esc}$. (b) The fractional change in these power spectra
relative to the model with no reionization.  The boost on large scales
and supression on small scales are due to reionization.
The values of $f_{\rm esc}=0.05, 0.1$ and $1.0$ correspond to optical
depths due to reionization of $\kappa_i=0.014, 0.017$ and $0.034$
respectively, for $\Omega_{\rm b}=0.02$. In panel (a), the power
spectrum is multiplied by the square of CMB temperature $T_{\rm
CMB}=2.726 K$. }
\label{Fig. 2}
\end{figure}

\begin{figure}
\plotone{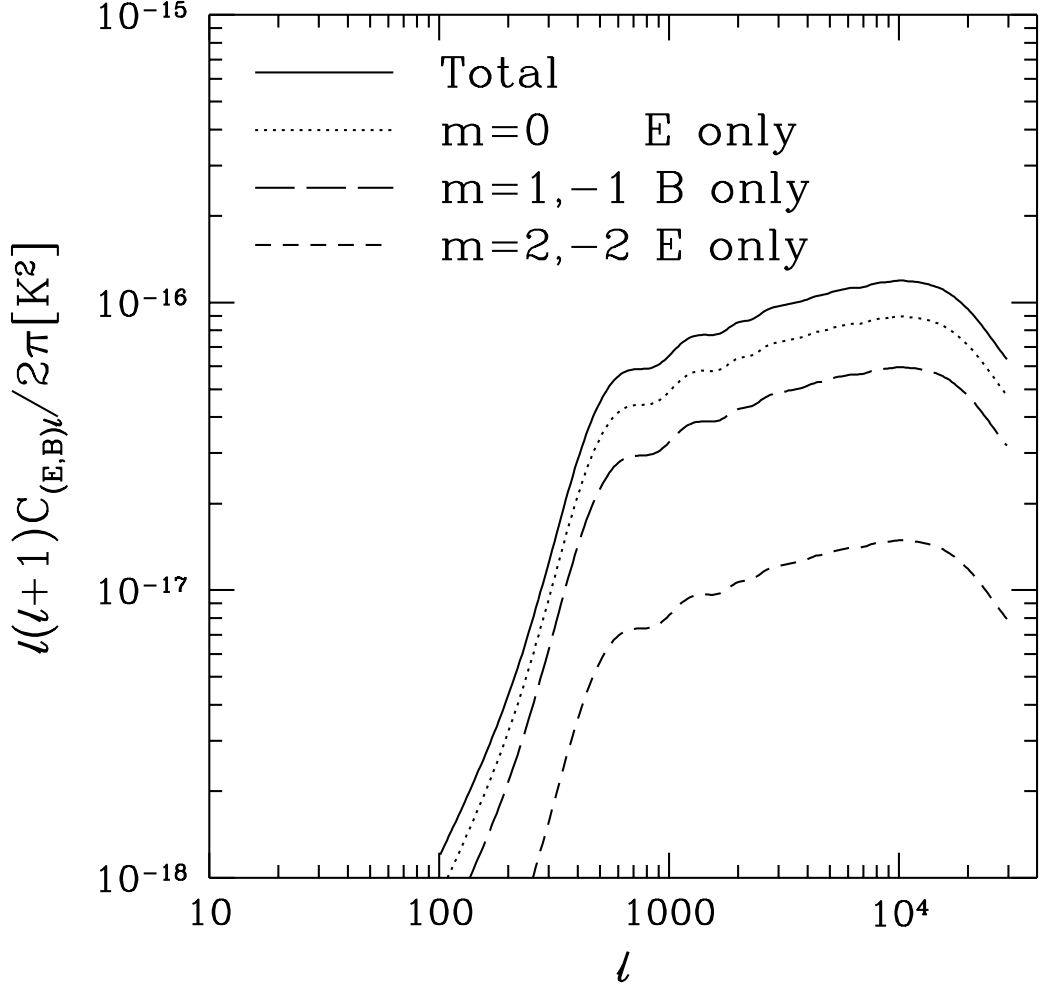}
\caption{ Contributions to the angular power spectrum of polarization
from the second-order effect from different values of $m$.  The curves
shown are computed using Model A with $f_{\rm esc}=1$. Note that the
long-dashed curve shows the contribution to the B-mode from only $m=1$
or $m=-1$ (not the sum of $m=1$ and $m=-1$), and similarly the
short-dashed curve shows the contribution to the E-mode from only
$m=2$ or $m=-2$. We find that the dominant contribution to the E-mode
is from $m=0$ and $m=\pm 2$, while for the B-mode it is from $m=\pm
1$.  There is no B-mode for $m=0$ and very small E- and B-modes for
$m=\pm 1$ and $m=\pm 2$, respectively.
The total power spectrum of the E-mode, plotted as the solid curve, 
and calculated by summing the $m=0,\pm 1$ and $\pm 2$
contributions, is the same as 
the total power spectrum of the B-mode. The
 power spectra are multiplied by $T^2_{\rm CMB}$. 
 }
\label{Fig. 3}

\end{figure}

\begin{figure}
\plotone{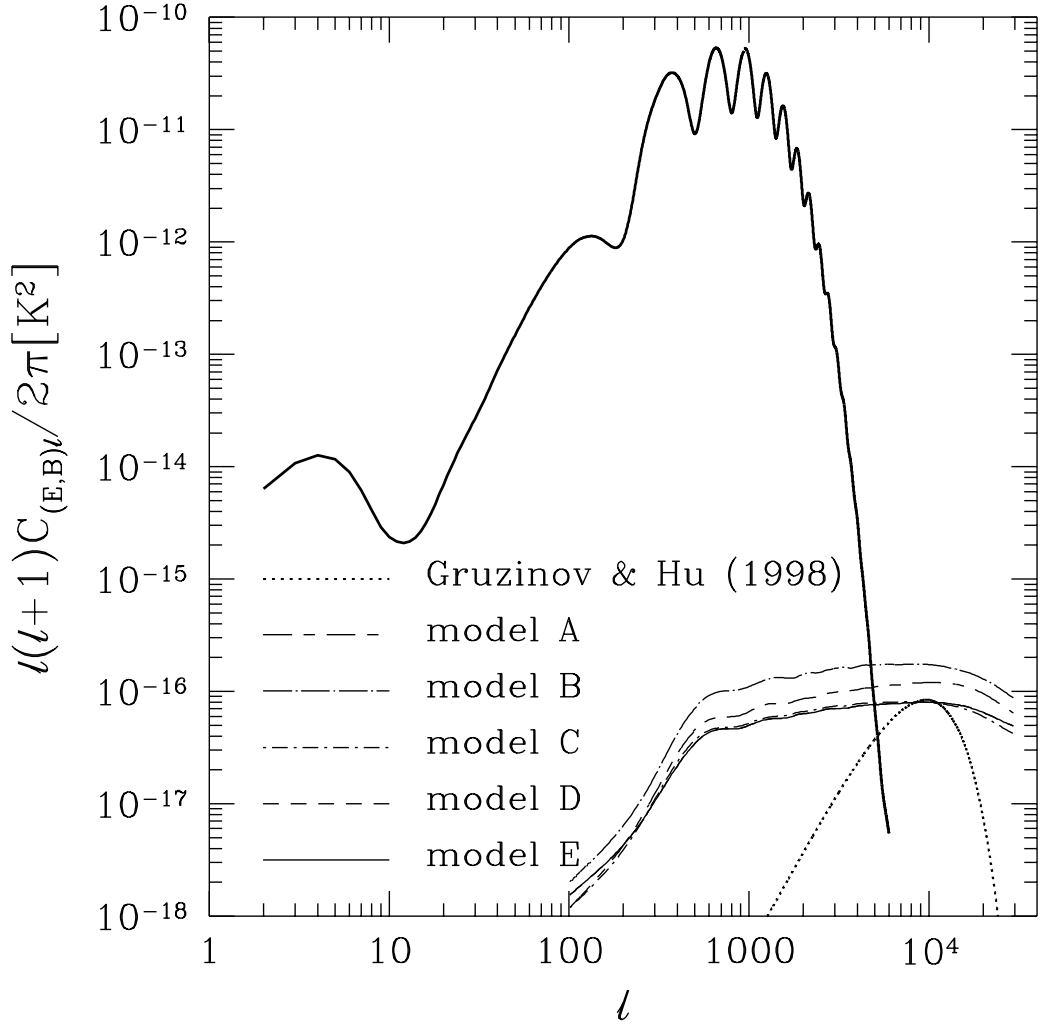}
\caption{
Power spectra of the second order effect for 
our models (note that the curve for model D is hidden under that for
model E). For comparison, the toy model of Gruzinov \& Hu (1998),
in which the ionized regions are 
randomly distributed, is also
shown. The heavy solid line shows the first-order anisotropy for
$\Omega_{\rm b}=0.02$ and $f_{\rm esc}=1.0$. Note that only $C_{El}$
contributes to the first-order effect, while in the second-order
effect $C_{El}$ 
and $C_{Bl}$ are equal.   } \label{Fig. 4}
\end{figure}

\begin{figure}
\plotone{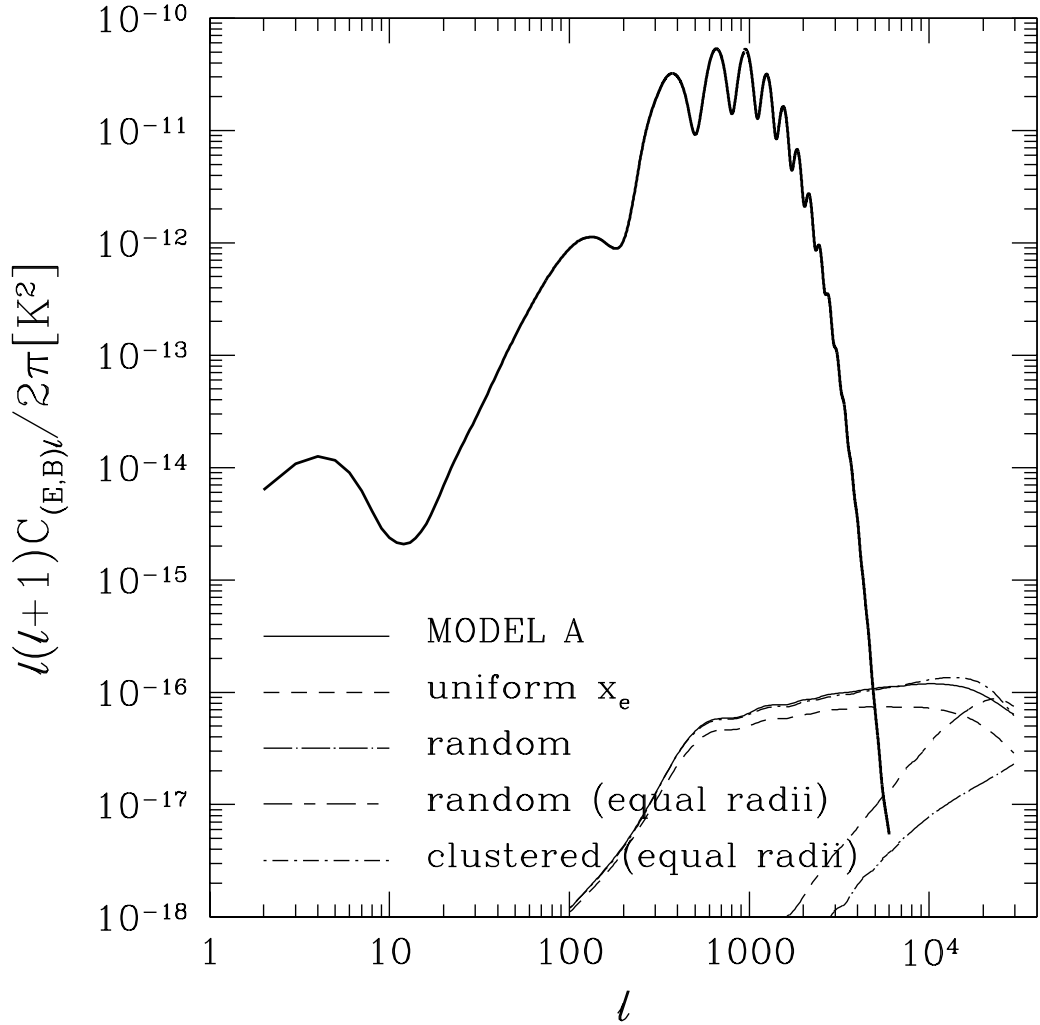}
\caption{ Effect on the second-order anisotropy of varying the assumed
geometry of the re-ionized regions. The models are all computed for
$\Omega_{\rm b}=0.02$ and $f_{\rm esc}=1.0$. The line for model~A is
repeated from Fig.~4. The line labelled ``uniform $x_e$'' is for the
case where the ionized fraction is uniform, and equal to the mean
value in model~A. For the line labelled ``random'', the total gas
density is forced to be uniform, and spheres are put down at random
positions but with the distribution of radii given by the galaxy
formation model. The case ``random (equal radii)'' is the same, except
that all spheres have the same comoving radius ($0.62 h^{-1}{\rm
Mpc}$). Finally, for the case ``clustered (equal radii)'', variations
in the total gas density are included, but all spheres have equal
radii at a given redshift, and are placed on random dark halos. See
the text for more details.}  \label{Fig. 5}
\end{figure}

\begin{figure}
\plotone{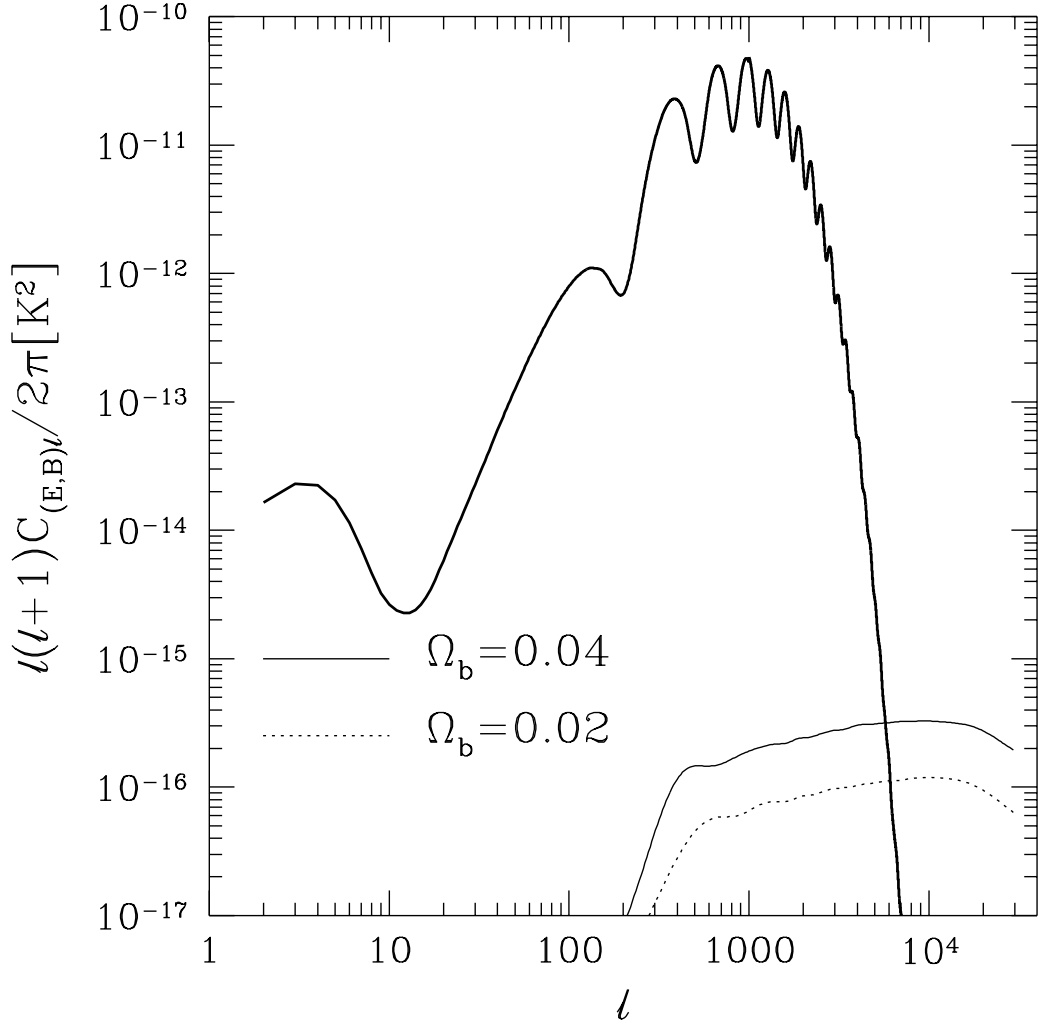}
\caption{
Effect on the secondary CMB polarization anisotropy spectrum
of varying $\Omega_{\rm b}$.  The curves shown are all computed using
Model A.  The solid line shows $\Omega_{\rm b}=0.04$ while the dotted
line shows $\Omega_{\rm b}=0.02$.  The heavy solid line shows the
first-order anisotropy for the model $\Omega_{\rm b}=0.04$ and $f_{\rm
esc}=1.0$.  The
 power spectrum is multiplied by $T^2_{\rm CMB}$.}
 \label{Fig. 6}
\end{figure}

\begin{figure}
\plotone{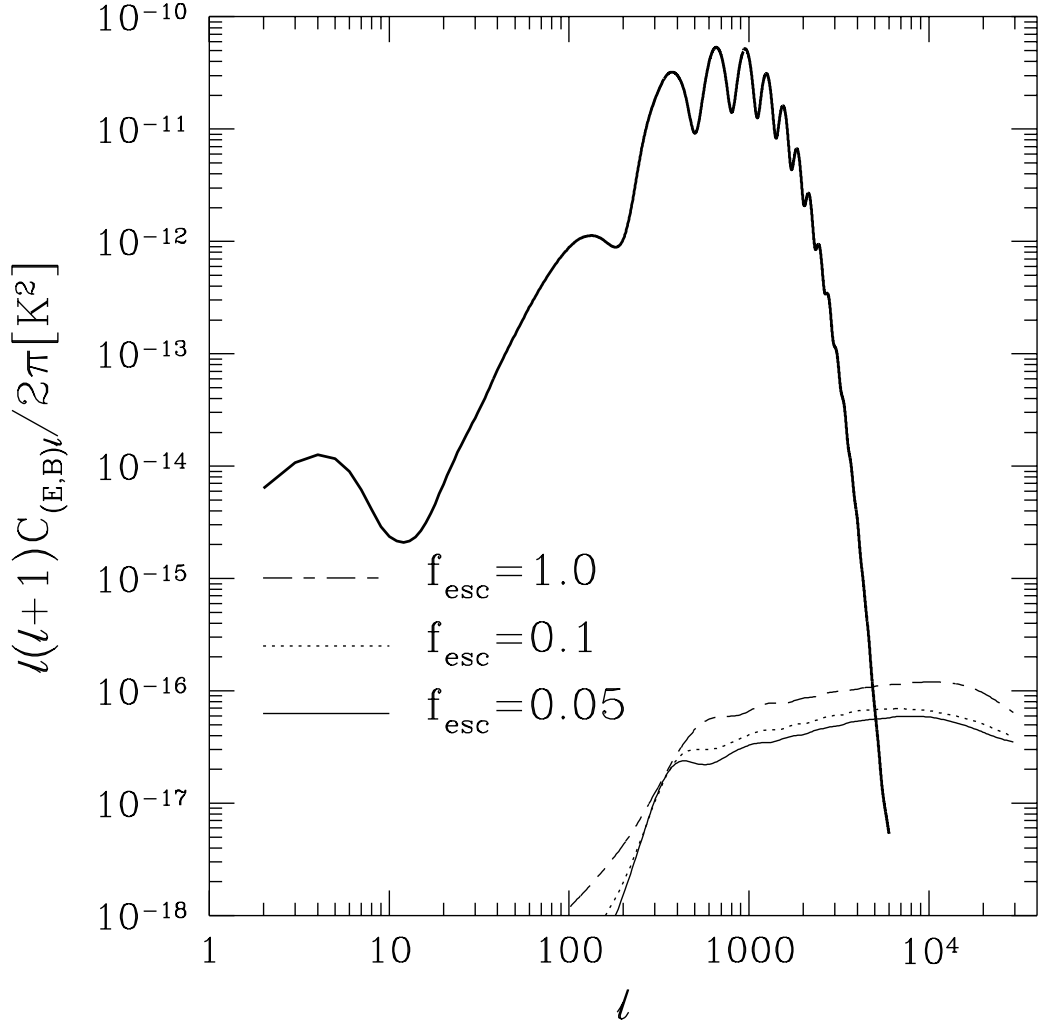}
\caption{Effect on the secondary CMB polarization anisotropy spectrum
of varying the escape fraction $f_{\rm esc}$.  The curves shown are
all computed using Model A and $\Omega_{\rm b}=0.02$.  Escape
fractions of $1.0, 0.1$ and $0.05$ are shown as indicated in the
figure.  The heavy solid line shows the primary anisotropy for the
model $\Omega_{\rm b}=0.02$. The
 power spectrum is multiplied by $T^2_{\rm CMB}$. }
 \label{Fig. 7}

\end{figure}

\end{document}